\documentclass [prb,aps,letterpaper,twocolumn,amsmath,amssymb,floatfix] {revtex4}

\usepackage{graphicx}
\usepackage{textcomp}
\usepackage{mathptmx} 

\newcommand{\dgc}{$^\circ\mathrm{C}$}

\begin{document}

\title{Coulomb blockade in a Si channel gated by an Al single-electron transistor}

\author{L. Sun}
\email{sunlyday@lps.umd.edu}
\author{K. R. Brown}
\altaffiliation{Present address: Time and Frequency Division, National Institute of Standards and Technology, 325 Broadway, Boulder, CO 80305}
\author{B. E. Kane}
\affiliation{Laboratory for Physical Sciences, 8050 Greenmead Drive, College Park, Maryland 20740, USA}

\begin{abstract}
We incorporate an Al-AlO$_x$-Al single-electron transistor as the gate
of a narrow ($\sim$100~nm) metal-oxide-semiconductor field-effect
transistor (MOSFET). Near the MOSFET channel conductance threshold, we
observe oscillations in the conductance associated with Coulomb
blockade in the channel, revealing the formation of a Si
single-electron transistor. Abrupt steps present in sweeps of the Al
transistor conductance versus gate voltage are correlated with
single-electron charging events in the Si transistor, and vice
versa. Analysis of these correlations using a simple electrostatic
model demonstrates that the two single-electron transistor islands are
closely aligned, with an inter-island capacitance approximately equal
to 1/3 of the total capacitance of the Si transistor island, indicating
that the Si transistor is strongly coupled to the Al transistor.
\end{abstract}

 \maketitle

 Single-electron transistors (SETs) are sensitive electrometers, able
to detect a small fraction of an electron charge, with a sensitivity
around $10^{-6}$~$e/\sqrt{\mathrm{Hz}}.$\cite{Aassime} They have been
extensively utilized for metrology\cite{Keller} and may have
applications in quantum information processing.\cite{Kane1,Buehler3}
While metal SETs are more common, Si SETs are desirable because of
their better stability\cite{Zimmerman1} and their ease of
incorporation into Si quantum computation
architectures.\cite{Emiroglu}

In this letter, we present results from a Si SET self-aligned and
vertically coupled to a metal SET. An Al-AlO$_x$-Al SET is
incorporated as the gate of a narrow ($\sim$100~nm) $\it{n}$-channel
metal-oxide-semiconductor field-effect transistor (MOSFET). Near the
channel conductance threshold, we observe Coulomb blockade
oscillations in the conductance, revealing the unintentional formation
of tunnel barriers in the channel and the creation of a Si SET. We
simultaneously monitor the conductance of each SET and observe a clear
correlation between single-electron charging events in the two SETs,
indicating that the SET islands are closely situated. The large charge
($\sim$0.33-0.35~$e$) induced onto the Al SET island by the addition
of another electron onto the Si SET island indicates that the Si SET
is strongly coupled to the Al SET, in the sense that the inter-island
capacitance is comparable to the total Si SET
capacitance. Consequently, this device structure could be used to
study individual tunneling events in the Si SET,\cite{Lu} and it
represents a single-electron current switch.\cite{Chan} It could also
be used to probe defect charge motion in MOS structures under large
electric field at low temperature,\cite{Brown} a potential source of
decoherence for Si quantum computation.\cite{Hu}

\begin{figure}[tb!]
\centering \includegraphics*[scale=0.4]{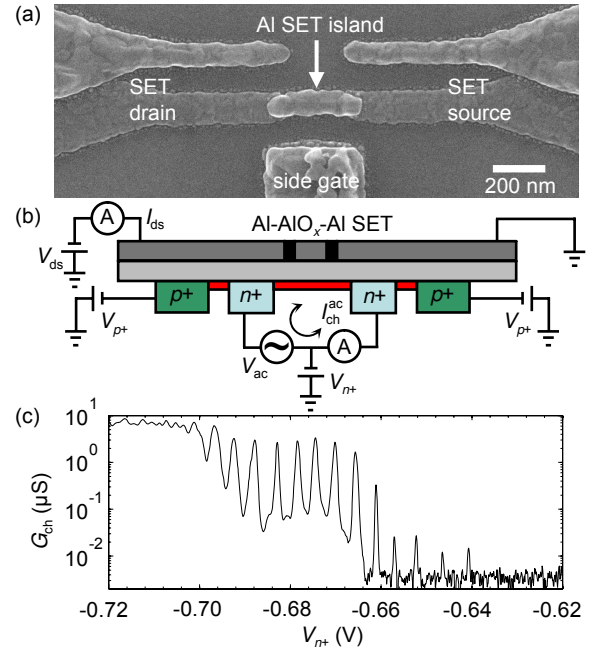}
\vspace{-10pt}
\caption{(a) SEM image of a typical device. The Al SET
island forms during the second of two evaporations at different
angles. (b) Schematic of the measurement circuit. The conductance of
each SET is measured using independent circuits. The red region
represents the MOSFET conducting channel confined between the two $\it{p}$+
regions. The circles containing the letter ``A'' represent current-sensitive amplifiers. (c) Coulomb blockade oscillations of the Si SET differential
conductance as a function of the relative bias $V_{n+}$ between the Al
SET and the Si SET at $V_g=-0.604$~V.}
\label{fig:SEM setup CBO}
\vspace{-15pt}
\end{figure}

Figure~\ref{fig:SEM setup CBO}(a) shows a scanning electron micrograph
of a typical sample. Fabrication starts with the oxidation of a
Si(100) wafer (resistivity $\rho > 8,000~\Omega$ cm) at 1000~\dgc,
yielding a SiO$_2$ thickness of about 20~nm. The wafer is selectively
ion implanted with P at an energy of 50~keV and an areal density of
$5\times 10^{14}$/cm$^2$ to create $\it{n}$+ contacts. To limit the extent
of the channel, $\it{p}$+ regions outside of the $\it{n}$+ contacts are created by
another ion implantation of B at 18~keV with an areal density of
$5\times10^{14}$/cm$^2$ [Fig.~\ref{fig:SEM setup CBO}(b)]. The
peak densities of both dopants are high enough to conduct at 20~mK and
reside close to the Si/SiO$_2$ interface. If the two implantations are
interchanged, a $\it{p}$-channel device can be made instead, so that both
polarities can be fabricated on a single chip. After both
implantations, the wafer is annealed at 950~\dgc\ for 60 s to
activate the dopants and to repair implantation damage. Electron-beam
lithography and self-aligned double angle evaporation are used to
fabricate the leads and island of the Al SET,\cite{Fulton} as well as
an Al side gate, used to modulate the conductance of both the Al SET
and the MOSFET channel [Fig.~\ref{fig:SEM setup CBO}(a)]. Finally,
the sample is annealed at 425~\dgc\ in forming gas for 30 min to
passivate dangling bonds at the Si/SiO$_2$ interface. 

All of the measurements that we present here were made on a single
device at a temperature of about 20~mK. A 1~T magnetic field was
applied to keep the Al SET in the normal state. The device
survived multiple thermal cycles to room temperature and displayed
only small background charge offset variations between cycles. To
avoid confusion, we present data from a single cooldown.
\begin{figure}[t]
  \centering \includegraphics*{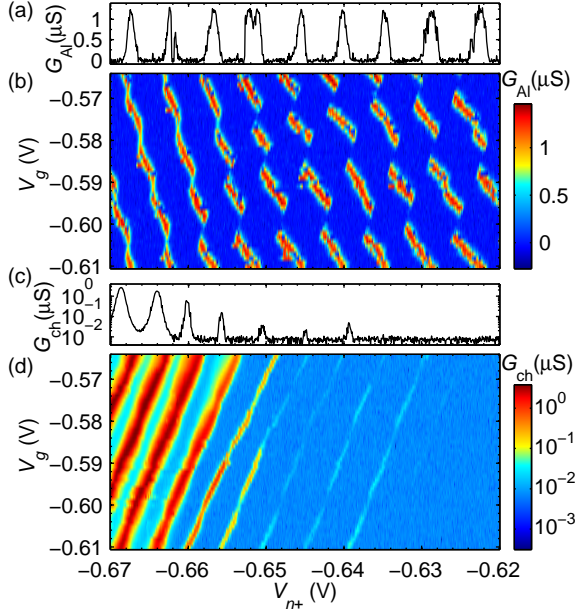}
\vspace{-10pt}
\caption{Simultaneously measured conductances of both SETs. [(a) and
(c)] Coulomb blockade oscillations of the Al and Si SET conductances,
respectively, at $V_g=-0.610$~V. [(b) and (d)] Conductance of the Al and the
Si SET, respectively, vs $V_g$ and $V_{n+}$.}
\label{fig:2D SET conductance}
\vspace{-13pt}
\end{figure}

Figure~\ref{fig:SEM setup CBO}(b) shows a schematic of the measurement
circuit. Both $\it{n}$+ contacts are dc biased at $V_{n+}$, while the Al SET
is grounded except for a small dc bias $V_{\mathrm{ds}}\sim$
100~$\mu$V. An ac excitation $V_{\mathrm{ac}}=10$~$\mu$V rms at 46~Hz
is applied between the two $\it{n}$+ contacts to measure the channel
differential conductance. The two $\it{p}$+ regions are dc biased at
potential $V_{p+}=-0.800$~V to confine the channel to a small region
between them. Figure~\ref{fig:SEM setup CBO}(c) presents typical data
for the channel differential conductance
($G_{\mathrm{ch}}=I^{\mathrm{ac}}_{\mathrm{ch}}/V_{\mathrm{ac}}$) as a
function of $V_{n+}$ at constant side gate voltage $V_g=-0.604$~V. The
appearance of Coulomb blockade oscillations is surprising, because no
tunnel barriers are deliberately engineered in the channel.

\begin{figure}[b]
\vspace{-8pt}
  \centering \includegraphics*{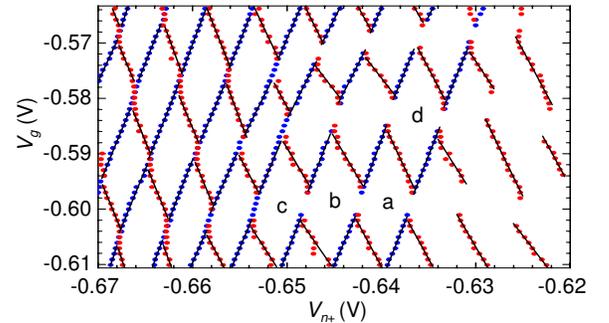}
\vspace{-10pt}
\caption{Conductance maxima of both SETs vs $V_{n+}$ and
$V_g$. Red dots and blue dots are Gaussian fits to the data in
Figs.~\ref{fig:2D SET conductance}(b) and \ref{fig:2D SET
conductance}(d), respectively. Black lines are a linear fit to the
points on each edge. The regions labeled a, b, c, and d are the four
hexagons whose parameters are presented in Table~\ref{T:hexagon
capacitances}.}
\label{fig:hexagon data}
\end{figure}

 In order to determine the coupling strength between the two SETs and
to infer the proximity of the islands, systematic sweeps of $V_g$ and
$V_{n+}$ are performed while the conductances of both SETs are
measured. Figure~\ref{fig:2D SET conductance}(b) shows the Al SET
conductance ($G_{\mathrm{Al}}=I_{\mathrm{ds}}/V_{\mathrm{ds}}$), and
Fig.~\ref{fig:2D SET conductance}(d) shows the channel SET
differential conductance versus $V_g$ and $V_{n+}$. If each SET were
electrostatically isolated from the other, its conductance maxima
would trace out straight lines in these graphs. Discontinuities from
this straight-line behavior evident in Figs.~\ref{fig:2D SET
conductance}(b) and \ref{fig:2D SET conductance}(d) are a signature of
discrete charging events close to the SET islands, events we would
expect to observe if the two islands were in close
proximity.\cite{Chan,Holleitner,Hubel} To confirm this hypothesis, the
maxima in Figs.~\ref{fig:2D SET conductance}(b) and \ref{fig:2D SET
conductance}(d) are fitted with Gaussians, and the resulting peak
centroids are plotted in Fig.~\ref{fig:hexagon data}. The two SETs
display a clear correlation: whenever an Al SET conductance peak trace
meets one from the Si SET, it makes an abrupt step, and vice
versa. Because each peak trace corresponds to a unit change in the
number of electrons on the corresponding SET island, this correlation
proves that single-electron charging events in one SET are coupled to
the other.

\vspace{-12pt}
To more quantitatively explain the above results, we have modeled the
device using the circuit depicted in Fig.~\ref{fig:electrostatic
model}(a). Under the assumption that the system can minimize its
electrostatic energy automatically by independently adjusting the
number of electrons $N_{\mathrm{Si}}$ and $N_{\mathrm{Al}}$ on the two
SET islands, the total electrostatic energy of this circuit is given
in matrix form by\cite{Conrad}
\vspace{-10pt}
\begin{equation}E(N_{\mathrm{Al}}, N_{\mathrm{Si}}, V_g, V_{n+})=\frac{1}{2}Q^T \left( \begin{array}{cc}
C_{\Sigma\_\mathrm{Al}}& -C_c\\ -C_c&
C_{\Sigma\_\mathrm{Si}}\end{array} \right)^{-1}Q.\end{equation} Here,
$C_{\Sigma\_\mathrm{Al}}=C_1+C_3+C_c+C_{g\_\mathrm{Al}}$ and
$C_{\Sigma\_\mathrm{Si}}=C_2+C_4+C_c+C_{g\_\mathrm{Si}}$ are the total
capacitance of the Al SET and of the Si SET island, respectively, and
$Q=(-eN_{\mathrm{Al}}+C_{g\_\mathrm{Al}}V_g+C_3V_{n+},-eN_{\mathrm{Si}}+C_{g\_\mathrm{Si}}V_g+C_2V_{n+})^T$
includes the virtual and actual charges on the SET
islands.\cite{Conrad} Under energy degenerate conditions, Coulomb
blockade is lifted, resulting in the maximal SET conductances. There
are in total six such degeneracy conditions associated with adding or
subtracting one electron from an SET island, determined by
$E(N_{\mathrm{Al}}, N_{\mathrm{Si}}, V_g,
V_{n+})=E(N_{\mathrm{Al}}$+$\delta N_{\mathrm{Al}},
N_{\mathrm{Si}}$+$\delta N_{\mathrm{Si}}, V_g, V_{n+})+(\delta
N_{\mathrm{Si}})eV_{n+}$. Here, $\delta N_{\mathrm{Al}}=0,\pm1$;
$\delta N_{\mathrm{Si}}=0,\pm1$; and $|\delta N_{\mathrm{Al}}$+$\delta
N_{\mathrm{Si}}|<2$, and $(\delta N_{\mathrm{Si}})eV_{n+}$ is the
extra work done by voltage source $V_{n+}$ when one electron tunnels
through junction $C_2$.\cite{Grabert} These equations establish the
hexagonal phase diagram depicted in Fig.~\ref{fig:electrostatic
model}(b), and there is a correspondence evident between this diagram
and the data in Fig.~\ref{fig:hexagon data}. However, the capacitances
associated with the Si SET appear to be bias voltage dependent,
resulting in the non-identical hexagons in the data.
\begin{figure}
\centering \includegraphics*{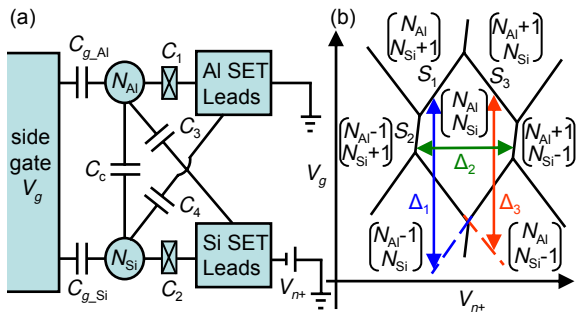}
\vspace{-10pt}
\caption{(a) Circuit model for the coupled SET system. $N_\mathrm{Al}$
and $N_\mathrm{Si}$ are the number of electrons on the Al and the Si SET
island, respectively. Due to the very small drain-source bias of each
SET, we can simplify the two tunnel barrier capacitances for each SET
to a single capacitance ($C_1$ and $C_2$) as shown. (b) Hexagonal
phase diagram based on the model in (a). Each hexagon represents a
configuration with a different number of charges on the SET
islands. $S_1$, $S_2$, and $S_3$ are the slopes of the hexagon
edges. $\Delta_1$, $\Delta_2$, and $\Delta_3$ are the separations
between opposite parallel edges of the hexagon.}
\label{fig:electrostatic model}
\vspace{-15pt}
\end{figure}

There are in total seven capacitance parameters in our circuit
model. $C_1=282\pm6$~aF is extracted from diamond chart measurements
(not shown) with the MOSFET in the off state ($V_{n+}=0$~V). $C_1$ is
dominated by overlap between the Al SET leads and its island and
should be insensitive to the presence or absence of an underlying
MOSFET channel. The remaining six parameters can be extracted from the
slopes $S_1$, $S_2$, $S_3$ and the separations $\Delta_1$, $\Delta_2$,
$\Delta_3$ of each hexagon in Fig.~\ref{fig:hexagon data}, as defined
in Fig.~\ref{fig:electrostatic model}(b). We extract these parameters
from the data as follows. First, the boundaries given by $\delta
N_{\mathrm{Al}}$+$\delta N_{\mathrm{Si}}=\pm1$ (the nominally straight
lines traced out by the data in Fig.~\ref{fig:hexagon data}) are each
fitted to a line. Boundaries corresponding to $\delta
N_{\mathrm{Al}}$+$\delta N_{\mathrm{Si}}=0$ (an effective transfer of
an electron from one island to the other) are not clearly visible, so
they are determined by neighboring intersections of the visible
boundaries. To compensate for gradual changes in the capacitances with
bias voltage, averages are made for the slopes and separations from
opposite boundaries within each hexagon. Then, the six unknown
capacitances can be extracted by solving six analytical equations
relating $S_1$, $S_2$, $S_3$, $\Delta_1$, $\Delta_2$, and $\Delta_3$.
\begin{table}[t]
\caption{Capacitances of the four hexagons labeled in
Fig.~\ref{fig:hexagon data} for the circuit model in
Fig.~\ref{fig:electrostatic model}.}
\begin{tabular*}{0.48\textwidth}{@{\extracolsep{\fill}}c c c c c c c}
\hline
\hline
hexagon&$C_2$(aF)&$C_c$(aF)&$C_3$(aF)&$C_4$(aF)&$C_{g\_\mathrm{Al}}$(aF)&$C_{g\_\mathrm{Si}}$(aF) \\
\hline
a&20$\pm$6&16$\pm$1&14$\pm$1&7$\pm$1&4.1$\pm$0.2&6.0$\pm$0.2\\
b&32$\pm$6&21$\pm$1&7$\pm$1&1$\pm$1&4.6$\pm$0.3&6.0$\pm$0.3\\
c&32$\pm$4&22$\pm$1&7$\pm$1&2$\pm$1&4.1$\pm$0.3&6.7$\pm$0.2\\
d&31$\pm$5&21$\pm$1&10$\pm$1&2$\pm$1&3.9$\pm$0.2&5.9$\pm$0.2\\
\hline
\hline
\end{tabular*}
\label{T:hexagon capacitances}
\vspace{-15pt}
\end{table}

\vspace{-10pt}
Discrete background charge motion near the SET islands, which changes
the electrostatics of the system, makes systematic study of all the
hexagons in Fig.~\ref{fig:hexagon data} difficult. The capacitances
for the four typical hexagons labeled in Fig.~\ref{fig:hexagon data}
are presented in Table~\ref{T:hexagon capacitances}. For hexagon a,
$C_{\Sigma\_\mathrm{Si}}=49\pm6$~aF. A diamond chart measurement
of the Si SET near this bias point (not shown) gives
$C_{\Sigma\_\mathrm{Si}}=49\pm3$~aF, in good agreement and
confirming the validity of the circuit model. Significantly, for all
the hexagons in Table~\ref{T:hexagon capacitances}, $C_c /
C_{\Sigma\_\mathrm{Si}} \sim$ 33\%--35\% indicates that the Si SET is
strongly coupled to the Al SET, while $C_c / C_{\Sigma\_\mathrm{Al}}
\sim$ 5\%--7\% explains why the discontinuities in Fig.~\ref{fig:2D SET
conductance}(d) are less obvious than those in Fig.~\ref{fig:2D
SET conductance}(b).

If the overlap between the two SET islands were perfect, the value of
$C_c$ as calculated from the Al SET island dimensions and the SiO$_2$
thickness would be about 30 aF. This is close to the values in
Table~\ref{T:hexagon capacitances} for hexagons b--d. The
small values for $C_4$ in these hexagons mean that there is almost no
overlap between the Si SET island and the Al SET leads. This strongly suggests
that the induced Si SET island is located directly beneath the Al SET
island.

We hypothesize the following reasons for the formation of an aligned
SET in the channel. Although the width of the SET island and that of the leads in
the evaporation shadow mask are the same, the SET island is formed
during the second evaporation. A slow pinch-off of features in the
mask during the first evaporation therefore makes the island slightly
narrower than the leads. If the angle between the evaporations is
incorrect, there may also be a lateral offset between the island and
leads. This island/leads width asymmetry and lateral offset may lead
to lateral constrictions in the MOSFET channel below, creating tunnel
barriers and therefore an SET in the channel aligned with the Al SET
above.

This SET sandwich architecture could be used to characterize the MOS
structure at low temperature via a cross-correlation measurement
between the two SETs.\cite{Buehler3} For example, measurements of the
Al SET show that about 10 electrons have already accumulated on the Si
SET island when the first measurable Si SET conductance peak
appears. This architecture could help identify sources of unwanted
charge motion that may also be sources of decoherence for Si quantum
computation. Because the SiO$_2$ layer could be made much thinner,
future experiments could more fully explore the strongly coupled
two-SET regime.

This work was supported by the Laboratory for Physical Sciences.

\end{document}